# Robust frequency stabilization and linewidth narrowing of a laser with large intermittent frequency jumps using an optical cavity and an atomic beam

**Won-Kyu Lee, Chang Yong Park, Myoung-Sun Heo, Dai-Hyuk Yu, and Huidong Kim***

*Center for Time and Frequency, Korea Research Institute of Standards and Science, Daejeon 34113, South Korea*
* khd250@kriss.re.kr

An experimental method is developed for the robust frequency stabilization using a high-finesse cavity when the laser exhibits large intermittent frequency jumps. This is accomplished by applying an additional slow feedback signal from Doppler-free fluorescence spectroscopy in an atomic beam with increased frequency locking range. As a result, a stable and narrow-linewidth 556 nm laser maintains the frequency lock status for more than a week, and contributes to more accurate evaluation of the Yb optical lattice clock. In addition, the reference optical cavity is supported at vibration-insensitive points without any vibration isolation table, making the laser setup more simple and compact.

## 1. Introduction

Frequency-stable and narrow-linewidth laser sources are essential tools for laser cooling and trapping of alkaline-earth-like atoms using intercombination transitions, which have various applications including quantum simulation [1], quantum computing [2], ultra-cold atomic or molecular experiments [3, 4], and optical clocks [5-8]. Among these, the second-stage magneto-optical trapping (MOT) of Yb atoms utilizes a narrow intercombination transition of $^1S_0 - {}^3P_1$ at 556 nm [5, 8]. The laser source for this MOT is usually obtained by a ring dye laser or more recently by using second harmonic generation (SHG) of an amplified output of either an Yb-doped fiber laser or an external cavity diode laser at 1112 nm. As the natural linewidth is very narrow (182 kHz) for this transition, the frequencies of the laser light sources have been narrowed and stabilized by various methods [4, 9-14]. Traditionally, this was accomplished by using a low expansion reference cavity [4], a transfer cavity [9], or a tunable optical cavity [10]. Alternatively, an atomic beam was used for a frequency reference adopting saturated absorption spectroscopy [11] or Doppler-free fluorescence spectroscopy [12]. The former method has an advantage of narrower linewidth, but the frequency drift due to the cavity spacer material ageing should be properly handled. The latter has little frequency drift, but the laser linewidth reduction is limited due to the residual Doppler broadening. Recently, the frequency of this 556 nm laser source has been stabilized using an optical frequency comb by phase locking [13] or frequency locking [14]. Although this method has been successfully demonstrated [13, 14], there are some drawbacks because an optical frequency comb is relatively costly, and the linewidth is limited by the short-term stability of a reference RF source for the optical frequency comb. Furthermore, robust long-term operation (continuous operation lasting longer than 1 day is preferred) is required in some applications of optical clocks, especially in the cases of optical clock comparisons [15], absolute frequency measurements [16], and dark matter detection [17].

Yb optical lattice clocks have been developed in our institute to prepare for more accurate future frequency standards [8, 12, 18, 19]. In the initial stage of the development, we used Doppler-free fluorescence spectroscopy in an atomic beam to stabilize the laser frequency for the second-stage MOT. This system has been upgraded by using a simple low expansion optical cavity as a frequency reference, and the number fluctuation of the atoms trapped in optical lattice has been significantly improved with this linewidth-narrowed laser source. However, the continuous operation of the lattice clock was limited to less than a few hours mainly due to a failure of frequency locking of a fiber laser for the second-stage MOT, which was caused by large intermittent frequency jumps (of several MHz magnitude) of the fiber laser at 1112 nm. To solve this problem, a fluorescence signal from an Yb atomic beam was used for a slow feedback, and continuous frequency locking was possible for more than a week. In Section 2 of this article, we will describe the characteristics of the frequency jumps, as well as the properties and the advantages of the simple setup of a reference optical cavity without a vibration isolation table. Robust stabilization of the narrow-linewidth second-stage MOT laser source will be described in Section 3. The result of the application to the MOT experiment will be given in Section 4, followed by a summary in Section 5.

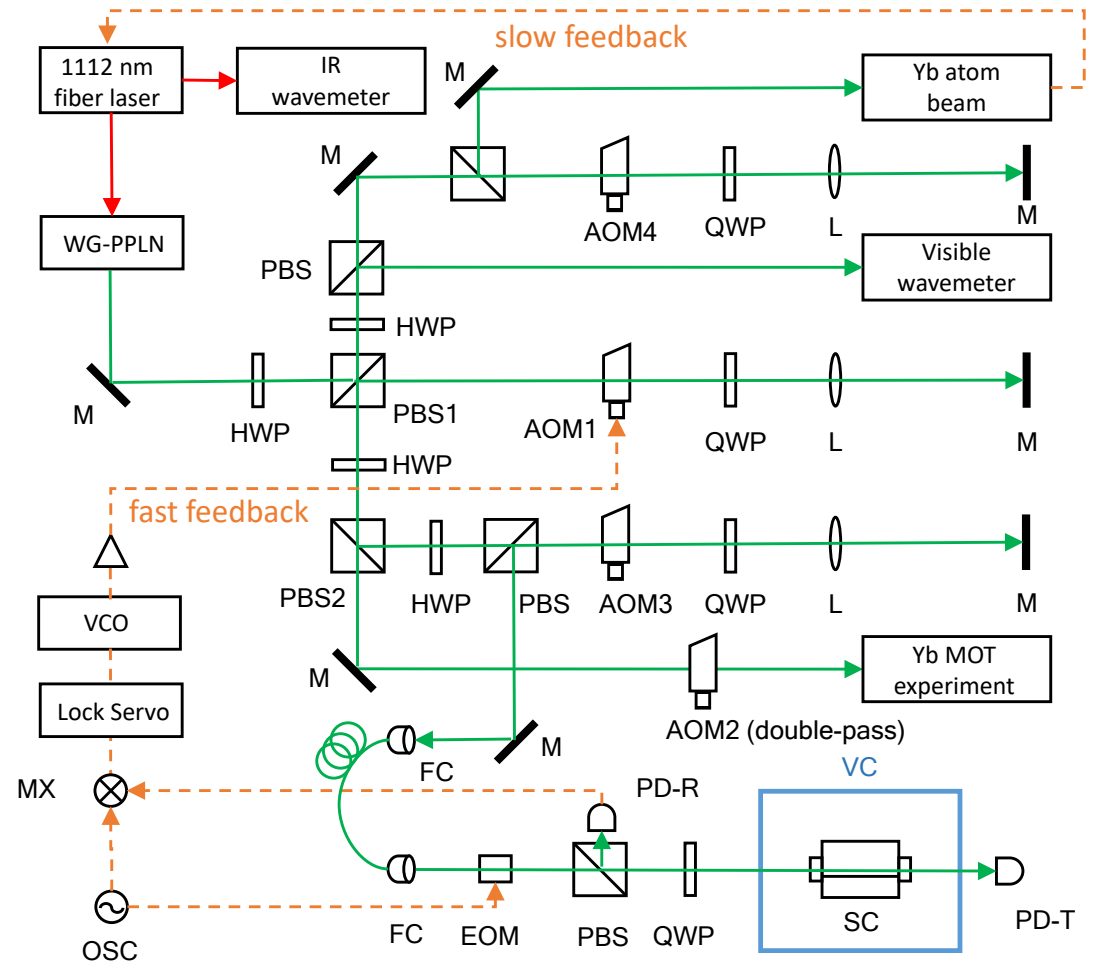

Fig. 1. Experimental setup for frequency stabilization of a fiber laser at 1112 nm. M, mirror; WG-PPLN, waveguide PPLN (periodically poled lithium niobate); HWP, half-wave plate; PBS, polarizing beam splitter; AOM, acousto-optic modulator; QWP, quarter-wave plate; L, lens; FC, fiber collimator; EOM, electro-optic modulator; PD-R, photodiode for reflection; PD-T, photodiode for transmission; SC, super-cavity; VC, vacuum chamber; OSC, two-channel function generator; MX, mixer; VCO, voltage controlled oscillator.

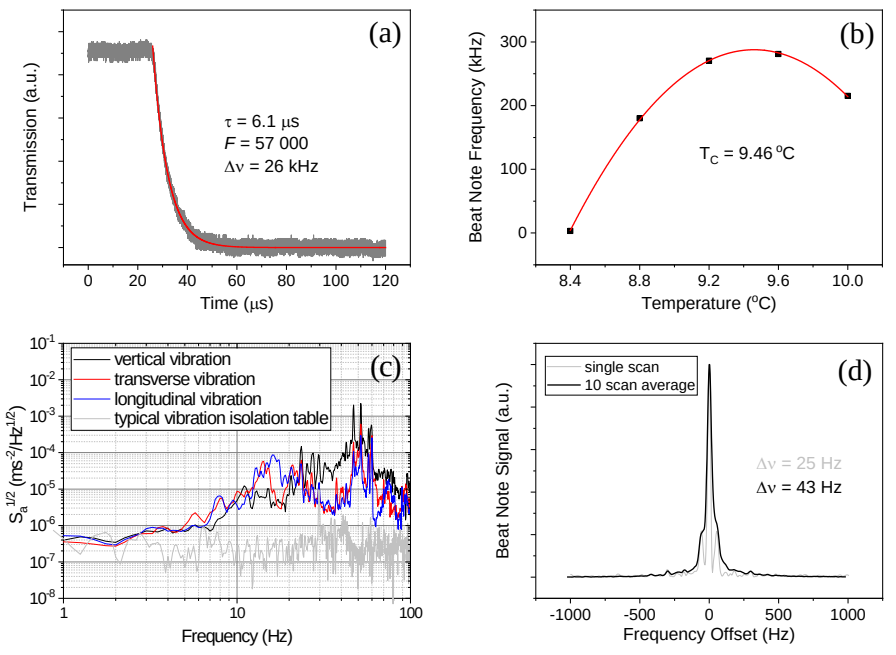

Fig. 2. (a) Cavity ring-down signal with the lifetime of 6.1 μs, which corresponds to a finesse of 57,000, (b) Experimental determination of the temperature for zero coefficient of thermal expansion of the ULE cavity, (c) Power spectral density of acceleration on an ordinary optical table in vertical, transverse, and longitudinal directions, and that on a typical vibration isolation table, (d) linewidth of a 578 nm laser, stabilized to the simple cavity setup used also for the 556 nm laser, obtained from the beat note with a second ultrastable 578 nm laser.

## 2. Simple high-finesse cavity setup for linewidth narrowing and the intermittent frequency lock failure

The experimental setup for the 556 nm laser source is shown in Fig. 1. We used a commercial single frequency Yb-doped fiber laser system at 1112 nm (500 mW of maximum output power after a fiber amplifier stage) to obtain a laser light source for the second-stage MOT in developing an Yb optical lattice clock [8]. The linewidth (for the measurement time of 120 μs) of the fiber laser was less than 100 kHz according to the manufacturer's specification. Output power of 90 mW at 556 nm was obtained by SHG using a fiber-coupled ridge-type waveguide periodically-poled lithium niobate (WG-PPLN) with an input power of 240 mW at 1112 nm. Previously we stabilized the frequency of the 556 nm laser to the $^1S_0(F = 1/2)$–$^3P_1(F = 3/2)$ transition of $^{171}$Yb atoms by using a fluorescence signal from an atomic beam [12]. The residual Doppler linewidth of the fluorescence signal was about 23 MHz, which was significantly larger than the natural linewidth of the narrow intercombination transition (182 kHz). As a result, there was a large fluctuation in the atom number trapped in the optical lattice due to both the short-term linewidth and mid-term frequency jitter. Thus, the system was upgraded so that the frequency of the 556 nm laser could be stabilized to a high-finesse optical cavity to obtain more stable operation of the MOT. The 556 nm output of the WG-PPLN double-passes through an acousto-optic modulator (AOM1), which acts as an actuator for the fast feedback for the frequency stabilization. Next, the laser light was split in two using a polarizing beam splitter (PBS2). The first part with higher laser power was sent to the Yb lattice clock after a double-pass AOM (AOM2), with which a broadband modulation and a frequency tuning was achieved for the optimum operation of the Yb optical lattice clock (the second-stage MOT and the state preparation). AOM2 also compensates the frequency difference between the cavity resonance and the Yb atom resonance. The other part went through another double-pass AOM (AOM3) and was coupled to a 2-m-long polarization maintaining single-mode fiber to be sent to a high-finesse optical cavity to stabilize the laser frequency to the resonance of the cavity by the Pound-Drever-Hall (PDH) method [20].

A horizontally-mounted cut-out type high-finesse ultra-low expansion (ULE) cavity [21] was used in this experiment. This cavity was installed on an ordinary optical table without any vibration isolation apparatus, but it was supported by Viton rubber pads at four points, which make the resonance frequency of the cavity vibration-insenitive [21]. The vertical, longitudinal, and transverse vibration sensitivities had been calculated by a finite element analysis to determine the optimum support position. This cavity was previously used for the frequency stabilization of a clock laser at 578 nm for an Yb optical lattice clock [12]. The finesse of the cavity was 161,000 at 578 nm (at the $^1S_0$ – $^3P_0$ clock transition wavelength), but as this would be used at a different wavelength, we newly measured the finesse at 556 nm to be 57,000 by the lifetime of 6.1 μs of the cavity ring-down signal as in Fig. 2(a). From this result the cavity linewidth was determined to be 26 kHz at 556 nm. We could cool down the cavity below the temperature of zero coefficient of thermal expansion (CTE) by using a Peltier cooler after we installed the cavity in a vacuum chamber made of aluminum as described in Ref. 21. We note that this had been impossible before with a stainless steel vacuum chamber, whose thermal conductivity was too small to cool down the cavity sufficiently. The temperature of zero CTE was determined to be 9.46°C as in Fig. 2(b) by a quadratic curve fitting. The temperature of the cavity is stabilized at the value for zero CTE within 10 mK. An ion pump (20 l/s) was used and the stabilized vacuum pressure was $9x10^{-7}$ Torr due to the outgassing from the aluminum chamber wall, however, this was sufficiently low for our application.

We could not directly measure the linewidth of the laser when this cavity was used at 556 nm, because there was no additional stabilized laser source at this wavelength to be compared with. However, the linewidth was conservatively estimated to be less than 1.3 kHz, considering that the cavity linewidth was 26 kHz at 556 nm and that the stabilized cavity transmission signal was maintained within less than 5% of its peak value at the resonance. To investigate the performance of the simple cavity setup more accurately, we stabilized the 578 nm clock laser using this setup without any vibration isolation apparatus. The linewidth measurement was done by a spectrum analyzer with the heterodyne beat note signal with another clock laser used in the Yb optical lattice clock experiment. Previously we could obtain a 1-Hz-level linewidth with this cavity on an active vibration isolation table [18], on which the power spectral density (PSD) of acceleration was at the level of $10^{-7}$ $ms^{-2}/Hz^{-1/2}$ (gray curve in Fig. 2(c)). However, when this cavity was placed on an optical table without vibration isolation, the vibration level was larger than that on a vibration isolation table by 100~1000 times (Fig. 2(c)). These acceleration measurement results were obtained by using a calibrated commercial vibration analyzer and a fast Fourier transform spectrum analyzer. Nevertheless, as is shown in Fig. 2(d), the laser linewidth could be narrowed to 25 Hz in a single scan with a resolution bandwidth of 18 Hz and a sweep time of 0.1 s. The laser linewidth was measured to be 43 Hz in 10 scan average, which includes mid-term frequency jitter. This result of narrow linewidth was possible thanks to the vibration-insensitiveness of the cavity support points. Although this linewidth measurement was done at 578 nm, the linewidth of the laser at 556 nm, which was narrowed by the same cavity setup, would be of the same order of magnitude because the finesse is smaller only by a factor of 3 at this wavelength.

The optical power incident on the cavity was about 50 μW. The cavity transmission was monitored by a photodiode (PD-T) and the reflected signal of the laser beam, which was phase modulated at 7 MHz by an electro-optic modulator (EOM), was detected by another fast photodiode (PD-R) and demodulated by a phase detector (MX), producing a PDH error signal for the fast feedback to AOM1 (the fast feedback bandwidth was limited to about 100 kHz due to the acoustic propagation delay in AOM1). At first, this PDH error signal was also used for the slow feedback to the PZT (piezoelectric transducer) of the fiber laser oscillator (the slow feedback bandwidth was about 400 Hz), unlike the case shown in Fig. 1, which was finally adopted to deal with large intermittent frequency jumps.

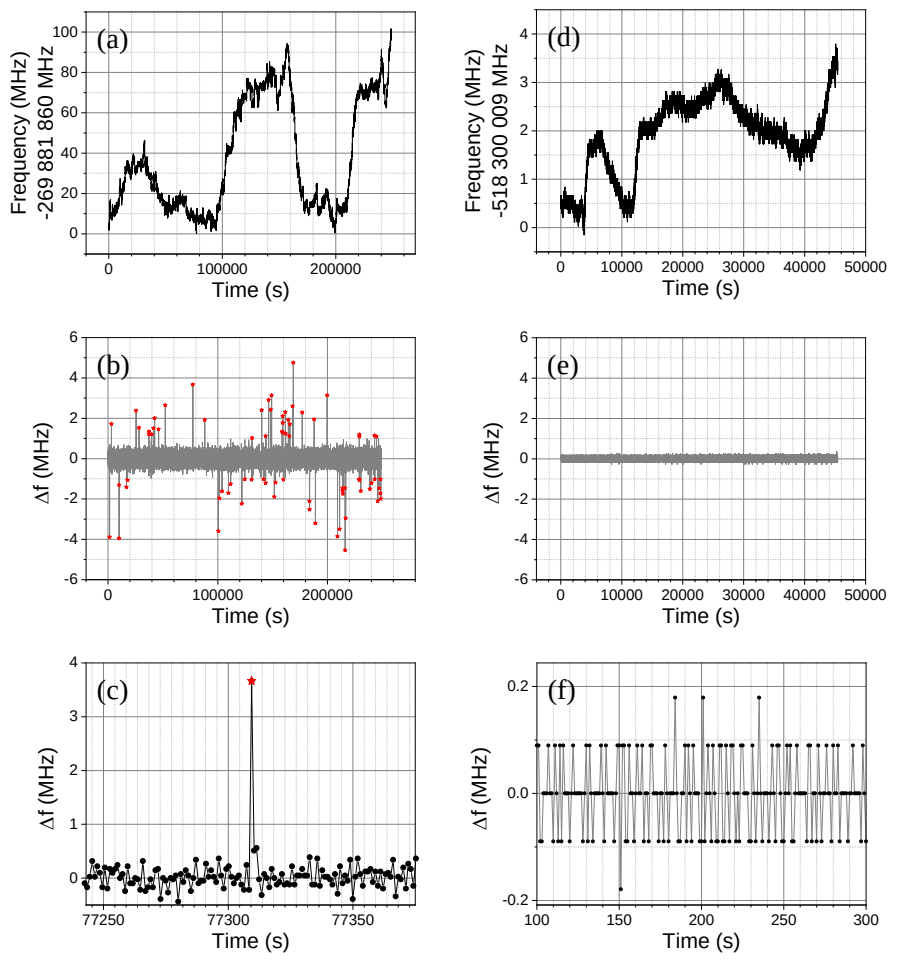

Fig. 3. (a) Long-term measurement of the frequency of a free-running 1112 nm fiber laser using a wavelength meter with time intervals of 1 s, (b) Difference of two adjacent frequency measurements of (a). Red stars represent frequency jumps larger than 1 MHz, (c) Magnified view of (b) around a typical frequency jump, (d) Long-term measurement of the frequency of a stable 578 nm laser using a wavelength meter with time intervals of 1 s, (e) Difference of two adjacent frequency measurements of (d), (f) Magnified view of (e).

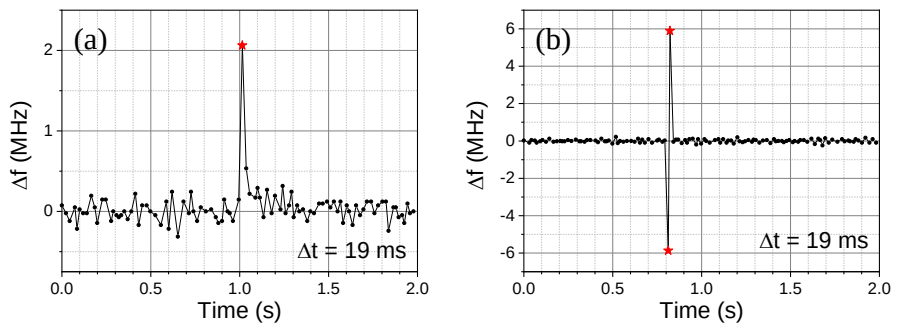

Fig. 4. (a), (b) Typical results of the difference of two adjacent short-term frequency measurement of a free-running 1112 nm fiber laser with time intervals of 19 ms. Red stars represent frequency jumps larger than 1 MHz.

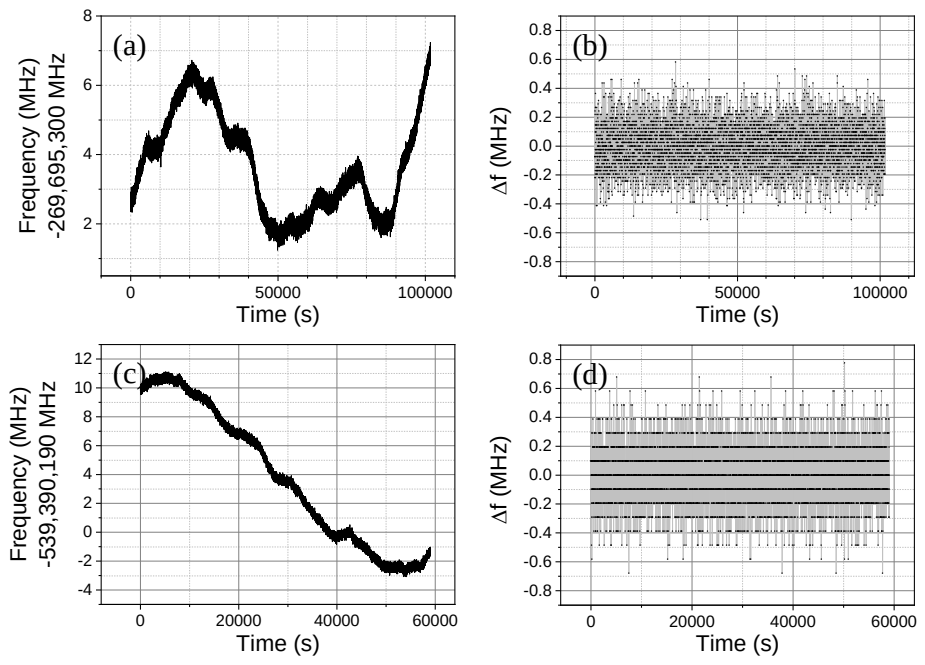

Fig. 5. (a) Long-term measurement of the frequency of a stabilized 1112 nm fiber laser using an IR wavelength meter with time intervals of 1 s, (b) Difference of two adjacent frequency measurements of (a), (c) Long-term measurement of the stabilized frequency of SHG of a fiber laser using a visible wavelength meter at 556 nm with time intervals of 1 s, (d) Difference of two adjacent frequency measurements of (c).

When we applied this narrow 556 nm laser system to the Yb clock experiment, we could find many experimental features had improved, including the atom number fluctuation trapped in the optical lattice. Full description of this improved features will be described in Section 4. However, a long-term continuous operation was not possible (limited to 2~3 hours in good condition) because of intermittent failures of the laser frequency locking at 556 nm. We attribute this frequency lock failure to frequency jumps in the 1112 nm fiber laser system. It has been known that the 1112 nm fiber laser can sometimes show large intermittent frequency excursions of tens of MHz [22]. Also, a multi-stage amplifier system should be used in Yb-doped fiber lasers at 1112 nm because of the reduced gain due to the large frequency offset from the gain peak, and small amount of optical feedback to the seed laser possibly causes large frequency excursions [23]. We suspect that these frequency jumps, which is larger than the locking range of the PDH lock, could make the frequency stabilization unrobust.

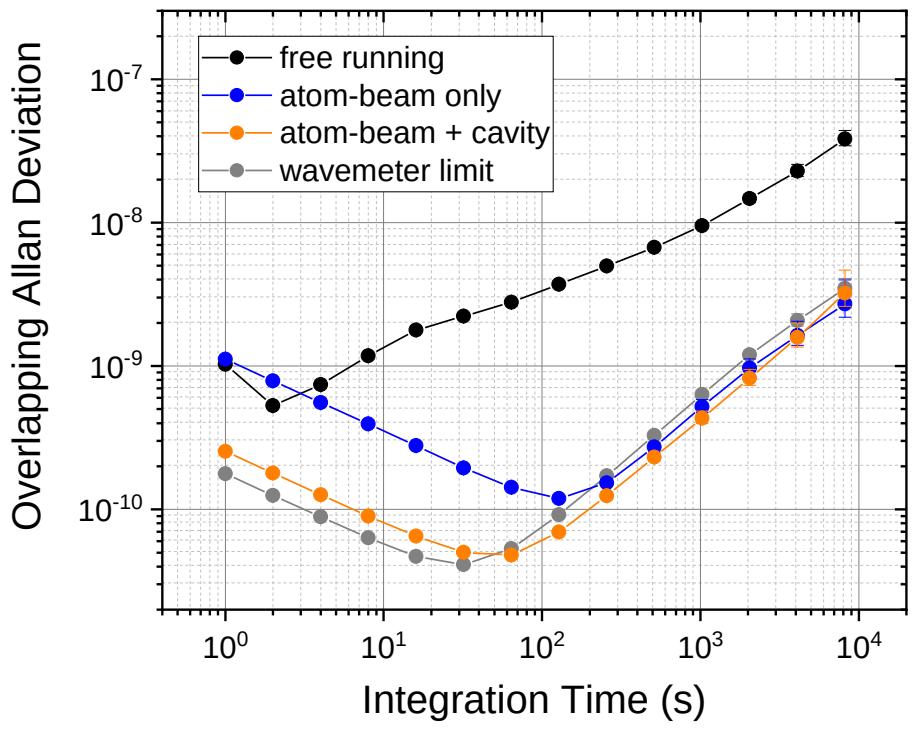

Fig. 6. Overlapping Allan deviations of the frequency of the laser, when it was free-running (black), stabilized by an atom beam only (blue), stabilized by both an atom beam and a high-finesse cavity (orange), and the wavelength meter limit (gray).

To investigate the presence of large frequency jumps, we measured the frequency of the free-running fiber laser at 1112 nm by using an IR high-resolution wavelength meter [24], and the result is shown in Fig. 3(a). The frequency data were acquired with time intervals of 1 s. The frequency difference Δf between two adjacent frequency measurements is shown in Fig. 3(b), and frequency jumps larger than 1 MHz are marked with red stars. As can be seen, 76 jumps larger than 1 MHz were observed in 245 500 s. When we magnify Fig. 3(b) around a frequency jump as in Fig. 3(c), we can see that a frequency jump is a fast event because every frequency jump is represented by a single point. To estimate the measurement limit of the wavelength meter, similar results were obtained using a stable laser at 578 nm, which was used in Yb optical clocks as a clock laser, by another wavelength meter of the same type but in visible wavelength range. The stable laser at 578 nm has a short-term linewidth of about 2 Hz and shows linear frequency drift of 0.056 Hz/s. As the frequency drifted only 5.6 kHz in $10^5$ s, the performance limit of the wavelength meter can be obtained by measuring the frequency of this laser. The results are shown in Figs. 3(d) ~ 3(f). The variation of the measured frequency value in Fig. 3(d) represents the long-term stability of the high-resolution wavelength meters caused by the ambient temperature change. Δf between two adjacent frequency measurement in Fig. 3(e) shows no frequency jump as expected for this stable clock laser. When we magnify Fig. 3(e), we can see in Fig. 3(f) digitized measurement values (with about 0.1 MHz steps) with very little fluctuation, which show the lower limit and the resolution of the high-resolution wavelength meters. It can be concluded from these results that the observed frequency jumps in Figs. 3(a) ~ 3(c) was not caused by the wavelength meter itself. Next, we changed the measurement rate of the wavelength meter to its maximum value. The time intervals for this maximum rate was calculated to be about 19 ms using the data number and the time elapsed. The frequency jumps were still represented by single points at this time scale, thus, it can be concluded that they are very fast phenomena (some typical results are shown in Fig. 4). It should be noted, however, we cannot say this frequency jumps are universal in 1112 nm Yb-doped fiber laser systems. We checked three fiber lasers at 1112 nm, which was available in our institute, however, only one system showed this feature of large intermittent frequency jumps.

## 3. Robust long-term frequency stabilization using atomic beam

To solve the problem of intermittent frequency lock failures of the 556 nm laser described in the previous section, we subsidiarily utilized an atomic spectroscopic signal obtained by an Yb atomic beam apparatus in addition to the PDH error signal using a high-finesse optical cavity. The oven was heated to 430°C to produce an Yb atomic beam. The atomic beam was collimated by using capillary tubes. By adapting a 6-way cross vacuum fitting, two viewports for the laser entrance and additional two viewports for fluorescence detection were installed perpendicularly to the atomic beam. As shown in Fig. 1, a small fraction of the 556 nm laser output from the WG-PPLN was split at the PBS1 before AOM1 where the fast feedback from PDH error signal is applied. This laser beam went through a double-pass AOM (AOM4), which compensates the frequency difference of the cavity resonance and the Yb atom resonance, and was subsequently retroreflected perpendicularly across the Yb atomic beam. The laser frequency was modulated also by AOM4 for the fluorescence spectroscopy with a frequency deviation of 2 MHz and with a modulation frequency of 29 kHz. The fluorescence signal from the atomic beam was collected by a lens and detected by a photomultiplier tube (PMT). The linewidth of this fluorescence signal was about 23 MHz due to residual Doppler broadening caused by the divergence of the atomic beam. The fluorescence signal from the PMT was demodulated to obtain an error signal, which was slowly fed back to the fiber laser PZT instead of the PDH error signal described in the previous section. The discrimination slope of this error signal was about 0.3 V/MHz and the locking range was very large (18 MHz), so that this error signal was expected to prevent the large frequency excursion of the fiber laser which would destroy the frequency lock to the optical cavity. We note that the linewidth of the fluorescence signal could be broadened more by a small tilt of the angle between the probe laser beam and the atomic beam, which would make the locking range even larger. When we applied this additional feedback for the frequency stabilization, the transmission signal of the optical cavity showed no difference, which means there was no degradation of linewidth narrowing by this additional feedback. With this change, the intermittent frequency jumps of the fiber laser were no more presented as shown in Fig. 5. In Fig. 5, we showed the long-term frequency measurement using both of the IR wavelength meter and the visible wavelength meter at 1112 nm (Fig. 5(a)) and 556 nm (Fig. 5(c)), respectively, when the fiber laser frequency was stabilized by both the optical cavity and the atomic beam spectroscopy. Also, as shown in Figs. 5(b) and 5(d), Δf between two adjacent frequency measurements shows no large frequency jump but only shows digitized measurement values with frequency steps which correspond to the resolution limit of each of the high-resolution wavelength meters. The frequency lock with this improvement was so robust, that it could maintain the frequency lock status even more than one week.

We could not precisely measure the frequency stability of the laser because there was no additional stabilized laser source at this wavelength. However, we could deduce some stability results by using a high-resolution visible wavelength meter at 556 nm within its limit as was done in Ref. 25. The results of the frequency instability in terms of Allan deviation are shown in Fig. 6. The wavelength meter limit was estimated by measuring the ultra-stable laser at 578 nm, which was used in Yb optical clocks as a clock laser, whose result is shown in Fig. 6 with gray dots. The frequency instability of the free-running fiber laser is shown with black dots. When we apply only the slow frequency lock by using the atomic beam, the short-term frequency instability (represented by blue dots) seems to be nearly the same as that of the free-running case, but it decreases with a slope of -1/2, which means the frequency lock was successful. After about 100 s, the instability increased with a slope +1 due to the thermal drift of the wavelength meter itself. Orange dots represents the frequency instability when both the optical cavity and the atomic beam were used for the frequency stabilization. Evidently there was an improvement in the short-term stability as we see the value at 1 s reaches the wavelength meter limit. Small difference between the orange and the gray dots can be attributed to the difference between the measured frequencies (556 nm and 578 nm). The frequency instability decreased also with a slope of -1/2 until it crosses the wavelength meter thermal drift limit around 30 s.

## 4. Application to MOT experiment

We successfully applied this robust, narrow-linewidth, and stable 556 nm laser system to the Yb optical lattice clock experiment, enabling more-than-1-day continuous clock operations. As mentioned in the previous section, the 556 nm laser frequency lock could be maintained for weeks. Meanwhile, the optical cavity resonance frequency has small linear drift because of the ULE material ageing, even though the cavity temperature was stabilized to the point with zero CTE, thus, the frequencies of AOM3 should be shifted as time passes. Considering the linear drift of 0.056 Hz/s, the accumulated frequency differences will be 4.8 kHz and 34 kHz for one day and for one week, respectively, but these frequency offsets are still much less than the natural linewidth of the atomic transition (182 kHz). Thus, we adjusted the central frequencies of AOM3 by maximizing the number of trapped atoms about once a week.

During this 1-week-long time interval, the temperature of the Yb atoms trapped in the optical lattice remained nearly the same, but the number of lattice-trapped atoms decreased slightly. The temperature of Yb atoms after the second-stage MOT was about (20 ± 4) μK with the adoption of the optical cavity for the frequency stabilization. The MOT temperature was not substantially improved, however, the position of the trapped atoms by the MOT was greatly stabilized, compared with that when we used only the atomic beam to stabilize the frequency, accordingly the relative atom number fluctuation in the optical lattice trap was significantly suppressed by a factor of ten, which made a more accurate frequency evaluation of the Yb optical lattice clock possible.

## 5. Summary

We discussed an experimental method for the robust frequency stabilization of a fiber laser at 1112 nm frequency-locked to a high-finesse optical cavity when the laser exhibits large intermittent frequency jumps. This was achieved by additional slow feedback to the laser using a fluorescence signal from an atomic beam with increased frequency locking range. The lattice-trapped atom number fluctuation was greatly improved compared with that when we used only the atomic beam for the frequency stabilization, which also contributed to a more accurate evaluation of the Yb optical lattice clock. Although these intermittent frequency jumps are not universal for all the Yb-doped fiber laser systems at 1112 nm, we believe this method would be useful in various situations because problematic frequency jumps can be induced not only by the faulty laser itself but also by various external perturbations. In addition, a simple setup for the reference cavity without any vibration isolation apparatus was developed using a cavity with optimized support positions for which the vibration sensitivity is minimal. This made the laser system for the second-stage MOT more compact.